\documentclass[10pt]{style/llncs}
\usepackage[cmex10]{amsmath}
\usepackage[noadjust]{cite}
\usepackage[pdftex]{graphicx}
\usepackage[caption=false,font=normalsize,labelfont=sf,textfont=sf]{subfig}

\usepackage{url}
\urldef{\mailamrabed}\path|amrabed@vt.edu|
\urldef{\mailtcc}\path|tcc@vt.edu|
\urldef{\maildslevy}\path|dslevy@mitre.org|
\newcommand{\keywords}[1]{\par\addvspace\baselineskip
\noindent\keywordname\enspace\ignorespaces#1}

\graphicspath{{figures/}}
\DeclareGraphicsExtensions{.eps,.pdf,.jpeg,.png}

\begin{document}
\pagenumbering{gobble}

\title{Intrusion Detection System for Applications using Linux Containers}

\author{Amr S. Abed\inst{1} 
\and Charles Clancy\inst{2}
\and David S. Levy\inst{3}}

\institute{Department of Electrical \& Computer Engineering, Virginia Tech, 
Blacksburg, VA\\
\mailamrabed\\
\and
Hume Center for National Security \& Technology, Virginia Tech, 
Arlington, VA\\
\mailtcc\\
\and
The MITRE Corporation, Annapolis Junction, MD\\
\maildslevy\\}

\maketitle
\begin{abstract}
Linux containers are gaining increasing traction in both individual and industrial use, and as these containers get integrated into mission-critical systems, real-time detection of malicious cyber attacks becomes a critical operational requirement. This paper introduces a real-time host-based intrusion detection system that can be used to passively detect malfeasance against applications within Linux containers running in a standalone or in a cloud multi-tenancy environment. The demonstrated intrusion detection system uses bags of system calls monitored from the host kernel for learning the behavior of an application running within a Linux container and determining anomalous container behavior. Performance of the approach using a database application was measured and results are discussed.
\keywords Intrusion Detection, Anomaly Detection, System Call Monitoring, Container Security, Security in Cloud Computing
\end{abstract}

\section{Introduction}
\label{sec_intro}
 Linux containers, such as Docker~\cite{docker} and LXC~\cite{lxc}, rely on the kernel namespaces and control groups (cgroups) for isolating the application running within the container. They provide a significantly more efficient alternative to virtual machines, since only the application and its dependencies need to be included in the container, and not the kernel and its processes. With the use of control groups and security profiles applied to containers, attack surface can be minimized~\cite{security}. However, attacks on mission-critical applications running within the container can still occur, and can represent an attack vector to the host kernel itself~\cite{security}. As a result, understanding when the container has been compromised is of key interest, yet little research has been conducted in this area.  

Indeed, Linux containers are typically used to run applications in a multi-tenancy cloud environments, where they share the same host kernel with other containers. In a multi-tenancy environment, the service provider is entitled by contractual means to monitor the behavior of containers running on the host kernel to provide safe environment for all hosted containers, and to protect the host kernel itself from the attack of a malicious container. However, providing information about the nature of the application running in the container, or altering the container for monitoring purposes is usually undesirable, and more often impermissible, especially when critical applications are running inside the container. Such constraints mandate the use of a host-based intrusion detection system (HIDS) that does not interfere with the container structure or application.

One source of attack originates from outside the host attacking the host kernel and/or the guest containers. Another source of attack comes from another containers residing on the same host and attacking neighboring containers. A third class of attack is when a container attacks the host kernel. To target these attacks, we propose a HIDS that monitors system calls between the container processes and the host kernel for malfeasance detection.

Utilizing system call traces for anomaly detection has been previously applied at the process level~\cite{forrest1996}\cite{hofmeyr1998}\cite{fuller2005}\cite{murtaza2013}, and has shown promising results when extended to the granularity of virtual machines (VMs)~\cite{mutz2006}\cite{alarifi2012}\cite{alarifi2013}. It has also been used to detect anomalies in Android applications by monitoring actions (aka system calls) included in their Android intents~\cite{ghorbanzadeh2014}.

There are two basic approaches to anomaly detection using system calls; sequence-based approach and frequency-based approach. The former approach keeps track of system call sequences in a database of normal behavior. The latter drops the order of the system calls while keeping the frequency of occurrence of each distinct system call. By not storing order information of the system call sequence, frequency-based techniques requires much less storage space while providing better performance and accuracy~\cite{fuller2005}. 

Bag of system calls (BoSC)~\cite{fuller2005} is a frequency-based approach that has been used as VM-based anomaly detectors in the past, and has been found to be a good performer~\cite{mutz2006}\cite{alarifi2012}\cite{alarifi2013}.  Particular advantages associated with the use of bags of systems calls, as opposed to sequences of system calls, are that it is computationally manageable~\cite{alarifi2012} and does not require limiting the application programming interfaces~\cite{mutz2006}.  

This paper serves to propose a real-time HIDS that can be used to passively detect anomalies of container behavior by using a technique similar to the one described in~\cite{alarifi2012}. We show that a frequency-based technique is sufficient for detecting abnormality in container behavior. The proposed system does not require any prior knowledge of the nature of the application inside the container, neither does it require any alteration to the container nor the host kernel, which makes it the first system to introduce opaque anomaly detection in containers, to the best of our knowledge. 

The rest of this paper is organized as follows. 
Section~\ref{sec_related} gives a brief summary of related work.
Section~\ref{sec_overview} provides an overview of the proposed system.
Section~\ref{sec_evaluation} discusses the system evaluation.
Section~\ref{sec_conclusion} concludes with summary and future work.
\section{Related Work}
\label{sec_related}
The \textit{Bag of System Calls} (BoSC) technique is a frequency-based anomaly detection technique, that was first introduced by Kang et al. in 2005~\cite{fuller2005}. In their paper, Kang et al. define the bag of system call as an ordered list $<c_1, c_2, \dots, c_n>$, where $n$ is the total number of distinct system calls, and $c_i$ is the number of occurrences of the system call, $s_i$, in the given input sequence. By applying different machine learning techniques, such as 1-class Na\"{i}ve Bayes classification and 2-means clustering, to the BoSC representation of two publicly-available system-call datasets, namely the University of New Mexico (UNM) dataset and the MIT Lincoln Lab dataset,  they were able to show that the BoSC has better performance and accuracy compared to STIDE~\cite{forrest1996}, one of the most famous and most popular sequence-based approaches.

 The \textit{Sequence Time-Delay Embedding} (STIDE) technique, introduced by Forrest and Longstaff~\cite{forrest1996}, defines normal behavior using a database of short sequences, each of size $k$. For building the database, they slide a window of size $k+1$ over the trace of system calls, and store the sequences of system calls. 
Although STIDE is a simple and efficient technique, it can be seen that by keeping the order information of the calls, the size of the database can grow linearly with the number of system calls in the trace. Some improvements to the STIDE technique were introduced in~\cite{hofmeyr1998} and~\cite{forrest1999}.

Another famous sequence-based intrusion detection technique is the one introduced in~\cite{lee1998}. The technique uses sliding windows (regions) of size $2l + 1$, with a sliding step of $l$, and relies on the RIPPER rule-induction application~\cite{cohen1995} to classify sequences of system calls into normal and abnormal regions. If the percentage of abnormal regions exceeds certain threshold, the trace is declared intrusive. 

A number of intrusion detection systems used sequences of system calls to train a Hidden Markov Model (HMM) classifier~\cite{forrest1999}\cite{wang2004}\cite{yeung2003}\cite{cho2003}\cite{hoang2003}. However, each system differs in the technique used for raising anomaly signal. Wang et al.~\cite{wang2004}, for example, raise anomaly signal when the probability of the whole sequence is below certain threshold. Warrender et al.~\cite{forrest1999}, on the other hand, declares a sequence as anomalous when the probability of one system call within a sequence is below the threshold. Cho and Park~\cite{cho2003} used HMM for modeling normal root privilege operations only. Hoang et al.~\cite{hoang2003} introduced a multi-layer detection technique that combines both outcomes from applying the Sliding Window approach and the HMM approach. 

Warrender et. al compared STIDE, RIPPER, and HMM-based methods in~\cite{forrest1999}. They concluded that all methods performed adequately, while HMM gave the best accuracy on average. However, it required higher computational resources and storage space, since it makes multiple passes through the training data, and stores significant amount of intermediate data, which is computationally expensive, especially for large traces. 

The \textit{Kernel State Modeling} (KSM) technique represents traces of system calls as states of Kernel modules~\cite{murtaza2013}. The technique observes three critical states, namely Kernel (KL), Memory Management (MM), and File System (FS) states. The technique then detects anomaly by calculating the probability of occurrences of the three observed states in each trace of system calls, and comparing the calculated probabilities against the probabilities of normal traces. Applied to Linux-based programs of the UNM dataset, the KSM technique shows higher detection rates and lower false positive rates, compared to STIDE and HMM-based techniques.
 
Alarifi and Wolthusen used system calls for implementing a HIDS for virtual machines residing in a multi-tenancy Infrastructure-as-a-service (IaaS) environment. They dealt with the VM as a single process, despite the numerous processes running inside it, and monitored system calls between the VM and the host operating system~\cite{alarifi2012}~\cite{alarifi2013}. 

In~\cite{alarifi2012}, they used the BoSC technique in combination with the sliding window technique for anomaly detection. In their technique, they read the input trace epoch by epoch. For each epoch, a sliding window of size $k$ moves over the system calls of each epoch, adding bags of system calls to the normal behavior database. The normal behavior database holds frequencies of bags of system calls. After building the normal-behavior database, i.e. training their classifier, an epoch is declared anomalous if the change in BoSC frequencies during that epoch exceeds certain threshold. For a sliding window of size $10$, their technique gave $100\%$ accuracy, with $100\%$ detection rate, and $0\%$ false positive rate. 

In~\cite{alarifi2013}, Alarifi and Wolthusen applied HMM for learning sequences of system calls for short-lived virtual machines. They based their decision on the conclusion from~\cite{forrest1999} that ``HMM almost always provides a high detection rate and a low minimum false positives but with high computational demand". Their HMM-based technique gave lower detection rates, yet required lower number of training samples. By using $780,000$ system calls for training, the resulting detection rate was $97\%$.

In their work, Chen et al.~\cite{ghorbanzadeh2014} applied HMM for recognizing malicious Android applications by monitoring actions (system calls) in Android intents issued by the application. They concluded that their technique, while capable of detecting malicious Android applications at runtime, did not have high performance, which they ascribed to not having enough Intent messages to further train the classifier. 
\section{Real-time Intrusion Detection}
\label{sec_overview}
\begin{figure}[!t]
\centering
\includegraphics[width=4in]{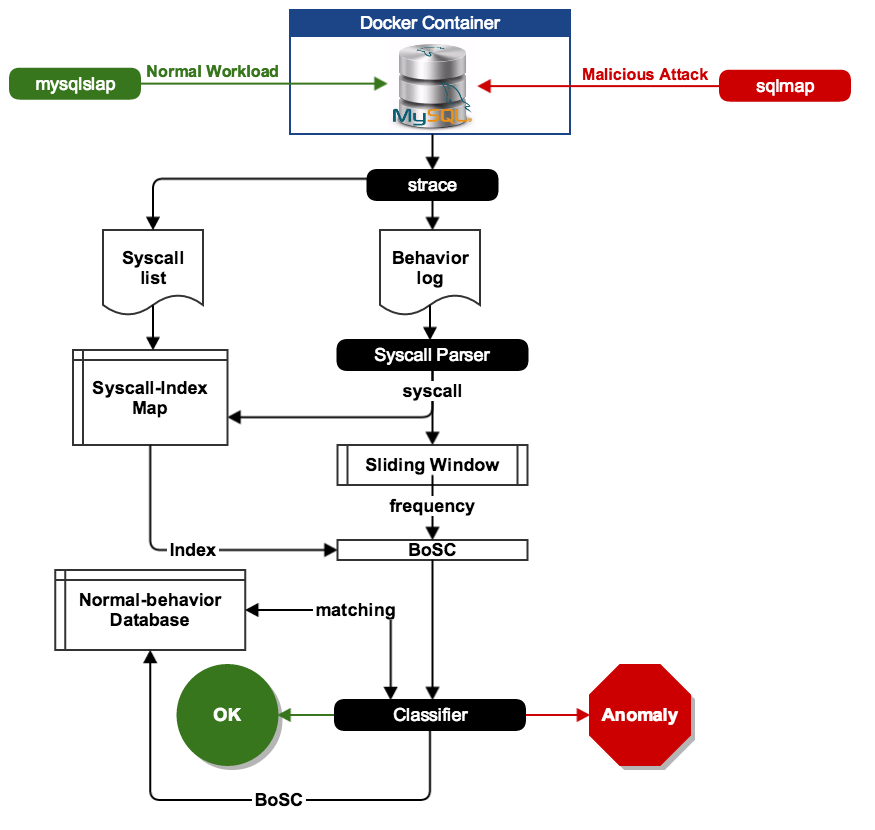}
\caption{Real-time Intrusion Detection System}
\label{fig_sys}
\end{figure}

In this paper, we propose a HIDS that uses a technique similar to the one described in~\cite{alarifi2012} to be applied to Linux containers. The technique combines the sliding window technique~\cite{forrest1996} with the bag of system calls technique~\cite{fuller2005}. The technique ignores the order of system calls, and only keeps track of the frequencies of the system calls in the current window. As described in section~\ref{sec_intro}, the system works in real time, i.e. it learns behavior of the container and detects anomaly at runtime. It also works in opaque mode, i.e. it does not require any prior knowledge about the nature of the container nor the enclosed application. Figure~\ref{fig_sys} gives an overview of the system architecture and data flow as described below. 

Our system employs a background service running on the host kernel to monitor system calls between any Docker containers and the host Kernel. Starting a new container on the host kernel triggers the service, which uses the Linux \texttt{strace} tool to trace all system calls issued by the container to the host kernel. The \texttt{strace} tool reports system calls with their originating process ID, arguments, and return values. 

In addition, \texttt{strace} is also used to generate a syscall-list file that holds a preassembled list of distinct system calls sorted by the number of occurrences. The list is collected from a container running the same application under no attack. The syscall-list file is used to create a syscall-index lookup table. Table~\ref{table_index} shows sample entries of a typical syscall-index lookup table. 
\begin{table}
\renewcommand{\arraystretch}{1.3}
\caption{Syscall-Index Lookup Table}
\label{table_index}
\centering
\begin{tabular}{cc}
Syscall & Index \\
\hline
select & 4\\
access & 12\\
lseek & 22\\
other & 40\\
\end{tabular}
\end{table}

The behavior file generated by \texttt{strace} is then parsed in either online or offline mode. In online mode, the system-call parser reads system calls from the same file as it is being written by the \texttt{strace} tool for real-time classification. Offline mode, on the other hand, is only used for system evaluation as described in section~\ref{sec_evaluation}. In offline mode, a copy of the original behavior file is used as input to the system to guarantee the coherence between the collected statistics. The system call parser reads one system call at a time by trimming off arguments, return values, and process IDs. 

The parsed system call is then used for updating a sliding window of size $10$, and counting the number of occurrences of each distinct system call in the current window, to create a new bag of system calls. As mentioned earlier, a bag of system calls is an array $<c_1, c_2, \dots, c_{n_s}>$ where $c_i$ is the number of occurrences of system call, $s_i$, in the current window, and $n_s$ is the total number of distinct system calls. When a new occurrence of a system call is encountered, the application retrieves the index of the system call from the syscall-index lookup table, and updates the corresponding index of the BoSC. For a window size of 10, the sum of all entries of the array equals $10$, i.e. $\sum_{i=1}^{n_s}{c_i}=10$. A sequence size of $6$ or $10$ is usually recommended when using sliding-window techniques for better performance~\cite{forrest1996}\cite{forrest1999}\cite{hofmeyr1998}. Here, we are using $10$ since it was already shown for a similar work that size $10$ gives better performance than size $6$ without dramatically affecting the efficiency of the algorithm~\cite{alarifi2012}.
Table~\ref{table_bosc} shows an example of this process for sequence size of $6$.
\begin{table}[!t]
\caption{Example of system call parsing}
\label{table_bosc}
\centering
\begin{tabular}{c|c|c|c}
Syscall & Index & Sliding window & BoSC\\
\hline
pwrite & 6 & [futex, futex, sendto, futex, sendto,  pwrite] & [2,0,3,0,0,0,\textbf{1},0,...,0]\\
sendto & 0 & [futex, sendto, futex, sendto, pwrite, sendto] & [\textbf{3},0,\textbf{2},0,0,0,1,0,...,0]\\
futex & 2 & [sendto, futex, sendto, pwrite, sendto, futex] & [3,0,\textbf{2},0,0,0,1,0,...,0]\\
sendto & 0 & [futex, sendto, pwrite, sendto, futex, sendto] & [\textbf{3},0,2,0,0,0,1,0,...,0]\\
\end{tabular}
\end{table}

The created BoSC is then passed to classifier, which works in one of two modes; training mode and detection mode. For training mode, the classifier simply adds the new BoSC to the normal-behavior database. If the current BoSC already exists in the normal-behavior database, its frequency is incremented by 1. Otherwise, the new BoSC is added to the database with initial frequency of $1$. The normal-behavior database is considered stable once all expected normal-behavior patterns are applied to the container. Table~\ref{table_db} shows sample entries of a normal-behavior database. 
 \begin{table}
\renewcommand{\arraystretch}{1.3}
\caption{Normal Behavior Database}
\label{table_db}
\centering
\begin{tabular}{cc}
BoSC & Frequency \\
\hline
0,1,0,2,0,0,0,0,1,0,3,0,1,0,0,0,1,0,0,1 & 15 \\
0,1,0,1,0,0,1,0,1,0,3,0,0,0,0,0,1,1,0,1 & 8 \\
0,1,0,2,0,0,5,0,0,0,0,0,0,0,0,0,1,0,0,1 & 2 \\
0,1,0,2,0,2,0,0,1,0,2,0,0,0,0,0,1,0,0,1 & 1 \\
\end{tabular}
\end{table}

For detection mode, the system reads the behavior file epoch by epoch. For each epoch, a sliding window is similarly used to check if the current BoSC is present in the database of normal behavior database. If a BoSC is not present in the database, a mismatch is declared. The trace is declared anomalous if the number of mismatches within one epoch exceeds a certain threshold.

Furthermore, a continuous training is applied during detection mode to further improve the false positive rate of the system. The bags of system calls seen during the current epoch are stored in a temporary current-epoch-change database rather than being added directly to the normal-behavior database. At the end of each epoch, if no anomaly signal was raised during the current epoch, the entries of the current-epoch-change database are committed to the normal-behavior database, to be included in classification for future epochs.
\section{System Evaluation}
\label{sec_evaluation}
\subsection{Environment setup}
For our experiments, we are using a Docker container running on a Ubuntu Server 14.04 host operating system. The docker image we used for creating the container is the official \texttt{mysql} Docker image, which is basically a Docker image with MySQL 5.6 installed on a Debian operating system. 

On container start, the container automatically creates a default database, adds users defined by the environment variables passed to the container, and then starts listening for connections. Docker maps the MySQL port from the container to some custom port on the host.

Since there is no dataset available that contains system calls collected from containers, we needed to create our own datasets for both normal and anomalous behavior. For that, we created a container from the \texttt{mysql} Docker image. A normal-behavior work load was initially applied to the container, before it got ``attacked" using a penetration testing tool. 
More details about generating datasets are given in sections~\ref{sec_normal} and ~\ref{sec_mal}.
\subsection{Generating normal workload}
\label{sec_normal}
For generating normal-behavior dataset, we used \texttt{mysqlslap}~\cite{mysqlslap}; a program that emulates client load for a MySQL server. The tool has command-line options that allow the user to select the level of concurrency, and the number of iterations to run the load test. In addition, it gives the user the option to customize the created database, e.g. by specifying the number of \texttt{varcher} and/or \texttt{int} columns to use when creating the database. Moreover, the user can select the number of insertions and queries to perform on the database. 

The tool runs on the host kernel, and communicates with the MySQL server running on the container. The values we used for generating the normal-behavior workload are shown in table~\ref{table_mysqlslap}.  
\begin{table}
\renewcommand{\arraystretch}{1.3}
\caption{Parameters used for automatic Load generation}
\label{table_mysqlslap}
\centering
\begin{tabular}{l l}
\textbf{Parameter} & \textbf{Value}\\
\hline
Number of generated \texttt{varchar} columns & 4\\
Number of generated \texttt{int} columns & 3\\
\hline
Number of simulated clients & 50 \\
Number of load-test iterations & 5\\
\hline
Number of unique insertion statements & 100\\
Total number of insertions per thread & 1000\\
Number of unique query statements & 100\\
Total number of queries per thread & 1000\\
\end{tabular}
\end{table}

Additionally, we used the SQL dump file of a real-life database to create schemas, tables, views, and to add entries to the tables, on the MySQL server of the container. 

\subsection{Simulating malicious behavior}
\label{sec_mal}
To simulate an attack on the container, we used \texttt{sqlmap}~\cite{sqlmap}; an automatic SQL injection tool normally used for penetration testing purposes. In our experiment, we are using it to generate malicious-behavior dataset by attacking the MySQL database created on the container. Similarly, the \texttt{sqlmap} tool runs on the host kernel, and communicates with the attacked database through the Docker proxy. 

We applied the following attacks on the container:
\begin{itemize}
\item Denial-of-Service (DoS) Attack: Using wild cards to slow down database. The attack generated an average of 37 mismatches
\item Operating system takeover attempt: Attempt to run Ôcat /etc/passwdÕ shell command (failed). Generated 279 mismatches
\item File-system access: Copy /etc/passwd to local machine. Generated 182 mismatches
\item Brute-force attack: We used the \texttt{--all} option of \texttt{sqlmap} to retrieve all info about the database management system (DBMS), including users, roles, schemas, passwords, tables, columns, and number of entries. The attack was strong enough to generate around 42,000 mismatches.
\end{itemize}
\subsection{Collecting container-behavior data}
A background service, running on the host kernel, automatically detects any newly started Docker container, and traces system calls of the new container using the Linux \texttt{strace} tool. 

The service relies on the Docker command \texttt{events} to signal the service whenever a new container is started on the host kernel. Upon detection of the new container, the service starts to trace all processes running,  on container start, within the control group (cgroup) of the container. The list of processes is retrieved from the \texttt{tasks} file located at \texttt{/sys/fs/cgroup/devices/docker/\$CID/tasks}, where \texttt{\$CID} is the long ID of the new container. The service also traces any forked child processes by using the \texttt{-F} option of the \texttt{strace} tool. 

To separate the normal behavior from the malicious behavior of the container for testing purposes, an indicator signal is injected into the behavior file before and after each attack, to be recognized by the classifier. 

\subsection{Training classifier}
We have implemented the classification system described in figure~\ref{fig_sys} in a Java application that uses the technique described in section~\ref{sec_overview}. 

The application starts by building a syscall-index hash map from the syscall-list file. The hash map stores distinct system calls as the key, and a corresponding index as the value. A system call that appears in the whole trace less than the total number of distinct system calls is stored in the map as ``other". Using ``other" for relatively rarely-used system calls saves space, memory, and computation time, as described in~\cite{alarifi2012}. By using a hash map, looking up the index of a system call is an $O(1)$ operation. 

The system call parser then reads one system call at a time from the behavior log file, and updates the normal-behavior database. The normal-behavior database is another hash map with the BoSC as the key and the frequency of the bag as the value. If the current bag already exists in the database, the frequency value is incremented. Otherwise, a new entry is added to the database. Again, by using a hash map for implementing the database, the time complexity for updating the database is $O(1)$.

As described in section~\ref{sec_overview}, the application uses the sliding window technique to read sequences of system calls from the trace file, with each sequence is of size 10. A bag of system calls is then created by counting the frequency of each distinct system call within the current window. The created bag of system calls is a frequency array of size $n_s$, where $n_s$ is the number of distinct system calls. When a new occurrence of a system call is encountered, the application retrieves the index of the system call from the syscall-index hash map, and the corresponding index of the frequency array is updated. The new BoSC is then added to the normal-behavior database.

\subsection{Classifier evaluation}
The generated normal-behavior database is then applied to the rest of the behavior file epoch by epoch for anomaly detection. For each epoch, the sliding window technique is similarly used to create BoSCs. The BoSCs noticed during the current epoch are added to temporary database. A mismatch is declared whenever a BoSC is not present in the database. If the number of mismatches exceeds a certain threshold, $T_d$, within one epoch, an anomaly signal is raised. Otherwise, the entries of the temporary database are committed to the normal-behavior database for future epochs. 

For evaluation purposes, the system-call parser recognizes the start-of-attack and end-of-attack signals injected during the data collection phase to mark the epochs involved in the attack as malicious. This information is used to accurately and automatically calculate  
the true positive rate (TPR) and false positive rate (FPR) metrics, defined as follows: 
 \begin{equation}
TPR=N_{tp}/N_{malicious}
\end{equation}
\begin{equation}
FPR=N_{fp}/N_{normal}
\end{equation}
where $N_{normal}$ and $N_{malicious}$ are the total number of normal and malicious sequences, respectively, and $N_{tp}$ and $N_{fp}$ are the number of true positives and false positives, respectively.

To evaluate the system accuracy with respect to different system parameters, we applied the classifier to the same input behavior file while varying the following test parameters:
\begin{itemize}
\item Epoch Size ($S$): The total number of system calls in one epoch. For our experiment, we used epoch size between 1000 and 10,000 with step of 500.
\item Detection Threshold ($T_d$): The number of detected mismatches per epoch before raising an anomaly signal. We used values between 10 to 100 with a step of 10 for each epoch size listed above.
\end{itemize}

\subsection{Evaluation Results}
We applied the proposed system to a trace of $3,804,000$ system calls, of which the classifier used $875,000$ system calls for training. The number of distinct system calls ($n_s$) was $40$, and the size of the normal behavior database was around $17k$ BoSCs. 

\begin{figure}
\centering
\includegraphics[width=5in]{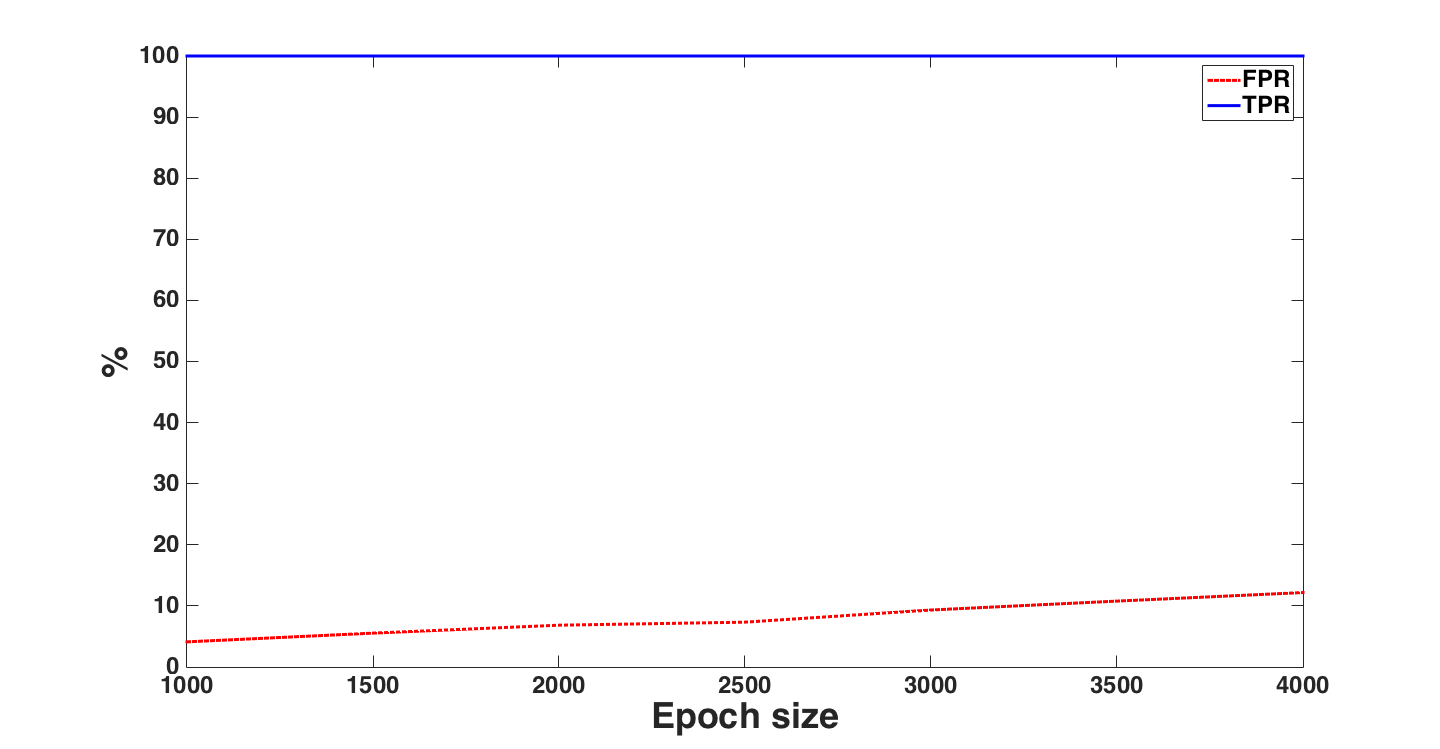}
\caption{Effect of changing epoch size on system accuracy}
\label{fig_10}
\end{figure}

The malicious data created a strong anomaly signal with an average of $695$ mismatches per epoch, as compared to an average of $33$ mismatches per epoch for normal data. For $S=1000$ and $T_d=10S$, the TPR is $100\%$ and the FPR is $2\%$. Figure~\ref{fig_10} shows the TPR and FPR of the system for different epoch sizes at the same detection threshold of $10$. It can be seen that the lower the epoch size, the lower the FPR. 

\begin{figure}
\centering
\includegraphics[width=5in]{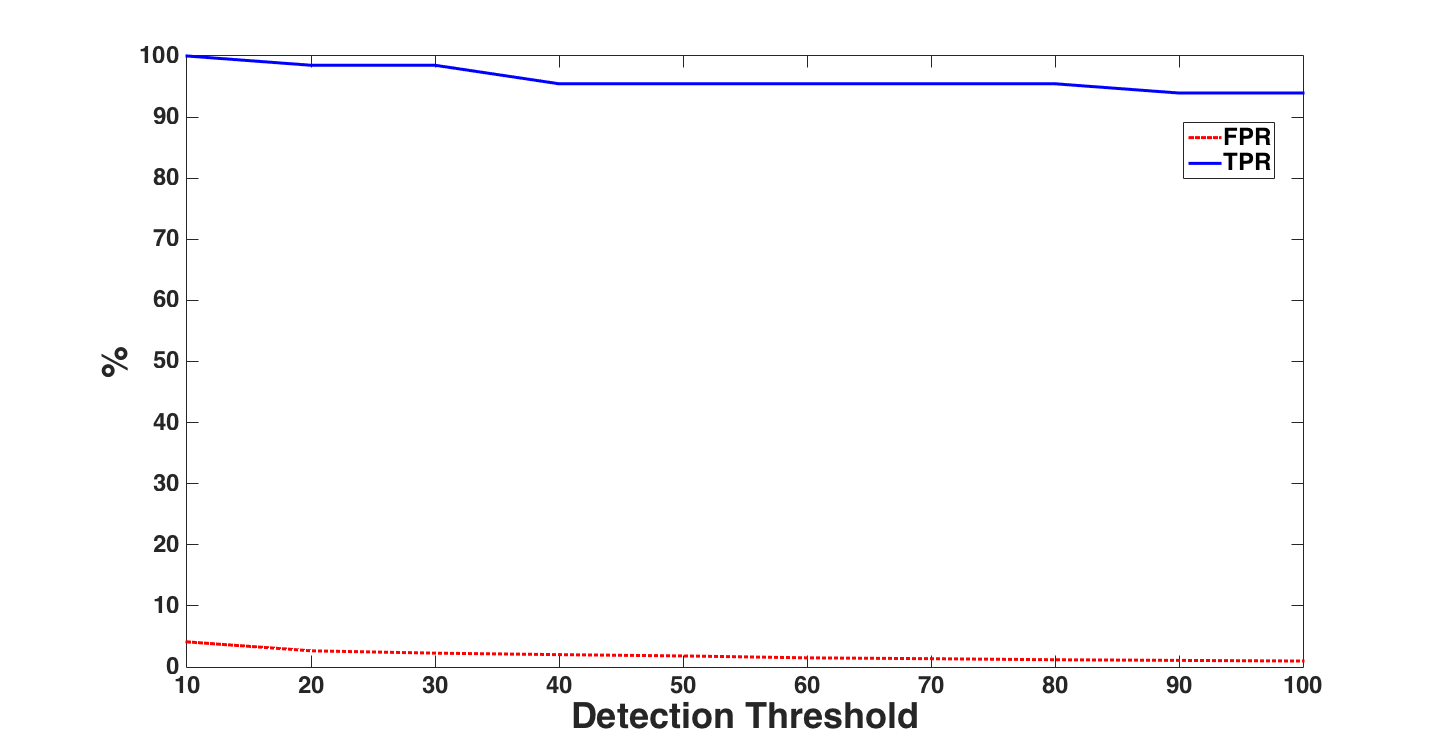}
\caption{Effect of changing detection threshold on system accuracy}
\label{fig_1000}
\end{figure}

As shown in figure~\ref{fig_1000}, the detection threshold highly affects the detection rate of the system especially when short-lived attacks are introduced to the container. 
\subsection{Complexity analysis}
By using hash map for the index map and the database, the time complexity for looking up an index for a given system call, and for updating the database with a new BoSC, are both $O(1)$ operations. The time complexity for comparing the database before and after an epoch $k$, and computing the similarity metric, is $O(n_k)$, where $n_k$ is the size of the database after epoch $k$. Hence, it can be seen that the algorithm used is linear in the size of the input trace. The time complexity of running an epoch of size $S$ is $O(S + n_k)$.

The algorithm only uses storage for the index map and the database. The index map holds \texttt{<String, Integer>} pairs. Assuming the average size of the system-call hash to be $8$ characters, the total size of an index map of size $n_s$ is $16n_s$ bytes (Typically $n_s<50$). The database stores array of bytes (a string) of size $n_s$ as the key, and an integer as the value. For a normal-behavior database of size $n_k$, the total size of the database is $(n_s + 8)n_k$ bytes. 
\section{Conclusion and Future Work}
\label{sec_conclusion}
In this paper, we have introduced an opaque real-time host-based intrusion detection system for detecting anomaly in the behavior of Linux containers. The proposed HIDS used a frequency-based anomaly detection technique previously applied to VMs. We were able to show that a high detection rate of $100\%$ is easily achievable using a low detection threshold of 10 mismatches per epoch. 

While the noticed FPR was relatively low (around $2\%$), we were not able to achieve a zero FPR for the used application and the applied learning technique. We attribute that to the non-repetitive behavior of the application, and the memory-based nature of the learning algorithm. It was noticed that applying the same workload to the MySQL database may not generate the exact same BoSCs, which is normally expected by an instance-based technique. Future work is to be directed to testing the system to applications of repetitive nature, such as a map-reduce application, and to modify the learning technique used to be more adaptive to slight changes of the BoSCs generated. 

Considering their popularity and simplicity of deployment, we are focusing on securing Docker containers for this research. However, the same methods can be extended to any other Linux containers, since they all share the same underlying architecture.

\subsubsection{Acknowledgments.} This work was funded by Northrop Grumman Corporation via a partnership agreement through S2ERC; an NSF Industry/University Cooperative Research Center.  
We would like to express our appreciation to Donald Steiner and Joshua Shapiro for their support and collaboration efforts in this work.

\bibliography{references}

\end{document}